# Micro-focused Brillouin light scattering study of the magnetization dynamics driven by Spin Hall effect in a transversely magnetized NiFe nanowire

M. Madami,[1] G. Gubbiotti,[2] T. Moriyama,[3] K. Tanaka,[3] G. Siracusano,[4] M. Carpentieri,[5] G. Finocchio,[4] S. Tacchi,[2] T. Ono,[3] G. Carlotti[1]

[1]*Dipartimento di Fisica e Geologia, Università di Perugia, Italy*
[2]*Istituto Officina dei Materiali del CNR (CNR-IOM), Unità di Perugia, c/o Dipartimento di Fisica e Geologia, Università di Perugia, Italy*
[3]*Institute for Chemical Research, Kyoto University, Japan*
[4]*Department of Electronic Engineering, Industrial Chemistry and Engineering University of Messina, Italy*
[5]*Department of Electrical and Information Engineering, Politecnico of Bari, Italy*

## Abstract

We employed micro-focused Brillouin light scattering to study the amplification of the thermal spin wave eigenmodes by means of a pure spin current, generated by the spin-Hall effect, in a transversely magnetized Pt(4nm)/NiFe(4nm)/SiO$_2$(5nm) layered nanowire with lateral dimensions 500×2750 nm$^2$. The frequency and the cross section of both the center (fundamental) and the edge spin wave modes have been measured as a function of the intensity of the injected dc electric current. The frequency of both modes exhibits a clear redshift while their cross section is greatly enhanced on increasing the intensity of the injected dc. A threshold-like behavior is observed for a value of the injected dc of 2.8 mA. Interestingly an additional mode, localized in the central part of the nanowire, appears at higher frequency on increasing the intensity of the injected dc above the threshold value. Micromagnetic simulations were used to quantitatively reproduce the experimental results and to investigate the complex non-linear dynamics induced by the spin-Hall effect, including the modification of the spatial profile of the spin wave modes and the appearance of the extra mode above the threshold.

## 1. Introduction

In the recent years the advent of pure-spin current generated by the spin-Hall effect (SHE) strongly stimulated the research in the field of spintronics, envisioning the way to realize devices which no longer require electric charge transfer. It has been demonstrated that when a spin current flows into a ferromagnetic material it exerts a torque on the magnetization, because of the spin-transfer torque (STT) effect, resulting in a local modification of the effective damping which can be either increased or reduced. [1,2,3] An increase of the local effective damping results in a larger suppression of thermally excited spin waves (SWs) which is potentially useful in reducing the thermal noise in nanoscale magnetic devices. On the other hand, a local reduction of the damping allows to amplify thermal SWs [4] and increase their propagation length of in a magnonic waveguide [5,6]. This offered the opportunity to envision new low-loss devices where the transmission and processing of signals can be effectively realized by SWs [7,8,9,10,11]. Complete compensation of damping and the consequent onset of auto-oscillations of the magnetization has been also recently demonstrated in magnetic nanostructures [12,13,14]. In general, at current larger than the compensation threshold, magnetic oscillations with large amplitude gives rise to nonlinear effects, for example nonlinear magnon-magnon interactions which can redistribute the energy in the spin wave spectrum of the magnetic system [4].

In the present work we investigated the amplification of thermal SWs by means of the STT due to a pure spin current, generated by SHE, in a Pt(4nm)/NiFe(4nm)/SiO$_2$(5nm) layered nanowire (NW) with lateral dimensions 500×2750 nm$^2$. By means of micro-focused Brillouin light scattering (micro-BLS) we were able to study the evolution of both the frequency and the cross section of the NW eigenmodes as a function of the intensity of the injected electric current ($i_{DC}$) flowing in the NW stack. We also observed the appearance of an additional mode, not visible in the thermal spectra, close to the $i_{DC}$ threshold. The experimental results are quantitatively reproduced and discussed by means of micromagnetic simulation.

## 2. Experimental results

The layer stack of Pt(4nm)/NiFe(4nm)/SiO$_2$(5nm) was prepared on a thermally oxidized Si wafer by magnetron sputtering. The film was patterned into 2750nm long and 500nm wide NW by electron beam lithography and Ar ion milling. Two Ti(5nm)/Au(30nm) electrodes were deposited on top of the NW edges leaving an optical access of 750nm×500nm in the center of the NW (Fig.1). Because of the different conductivities [15] and thicknesses of the Au electrodes and the Pt and Nife layers, the majority of the dc current flows in the Au electrodes up to the central aperture and then into the Pt wire, generating a spin current due to the SHE in Pt.

The micro-BLS measurements have been performed applying an external magnetic field at an angle of about 20° from the sample normal, because of physical constraints in the experimental apparatus which prevents to apply a fully in-plane magnetic field. The in-plane component of the external field is $H_{IP}$=60 mT and is directed along the *y* direction, i.e. perpendicularly to the NW as shown in Figure 1. We need to stress that the intensity of the external field (H=178 mT) is not sufficient to significantly tilt the NiFe magnetization out of the plane, so that the system is effectively in-plane magnetized. Micromagnetic simulations estimate an angle of about 7° out of the device plane for the magnetization of the NiFe NW. Since the direct current flows parallel to the NW, the three vectors $\mathbf{i_{DC}}$, $\mathbf{H_{IP}}$, and the normal to the film form a right-hand set, as needed in order to obtain an enhancement of the SWs in the magnetic material [12]. The magnetization dynamics in the NiFe NW was experimentally measured, at room temperature, by means of a micro-BLS setup described elsewhere [16]. About 5 mW of monochromatic light from a diode-pumped solid state laser (λ=532 nm) was focused, at normal incidence, on the NiFe surface, using a microscope objective, into a spot of about 300 nm of diameter. Light inelastically scattered from SWs in the NiFe NW was collected by the same objective and sent into a multi-pass (3+3) tandem Fabry-Perot interferometer for frequency analysis [17]. We only measured the anti-Stokes side of each spectrum in order to halve the total acquisition time. Anyway since measurements were performed at normal incidence and the thickness of the NiFe is only 4 nm the two sides of the spectrum are expected to be symmetric [18].

As a preliminary step of our experiment we measured the spectrum of thermally excited SWs in the NiFe NW, setting the intensity of the injected $i_{DC}$=0 mA. We measured micro-BLS spectra focusing the light in different points across the NiFe NW in order to detect SW eigenmodes with different spatial localization. In the top-left inset of Fig.2 we report two micro-BLS spectra measured, respectively, in the center of the NW (green spectrum) and close to the edge of the NW perpendicular to the applied field (red spectrum). In the former case, only one peak is present in the spectrum at about 5.8 GHz (M2), corresponding to the fundamental mode of the NW. This peak is also visible in the spectrum measured near the edge of the NW, due to the relatively large area occupied by the fundamental mode and by the finite spatial resolution of our technique. However, one can see also a second peak at about 4.3 GHz (M1), which corresponds to the edge-mode, i.e. a mode localized in the region where the effective field is reduced due to the demagnetizing field of the transversely magnetized NW [19,20,21].

In the next step of our experiment the laser spot was focused in an intermediate position between the edge and the center of the NW in order to study the evolution of both modes as a function of the intensity of the injected current $i_{DC}$. Looking at the sequence of spectra in Fig.2 it is clear that on increasing $i_{DC}$ both the edge (M1) and the fundamental mode (M2) intensities are enhanced, while their frequencies are red-shifted with respect to their values at $i_{DC}$=0 [12]. However, the above effects proceed rather slowly as $i_{DC}$ is varied from zero to about 2.5 mA, while both intensity enhancement and frequency decrease become more substantial as $i_{DC}$ is increased up to 2.8-3.0 mA where the intensity saturates. A graph of the inverse of the integrated intensity (top-right inset of Fig.2) clearly show a linear behavior from which we extrapolate a threshold value for $i_{DC}$ of about 2.8 mA. Looking more in detail into the peaks structure one can notice that the edge mode peak (M1) becomes narrower, due the reduced effective damping as $i_{DC}$ is increased, while maintaining a symmetric shape. On the other hand the shape of the fundamental mode (M2) is slightly asymmetric with the rising slope being sharper than the trailing one and this feature remains until the direct current reaches a value of about $i_{DC}$=2.8 mA. We speculate that the observed asymmetric shape of the fundamental mode peak is due to a convolution of different eigenmodes of the NW spectrum. In this set of eigenmodes the real fundamental mode is the one with the largest intensity and the lower frequency and is responsible for the sharpness of the peak rising slope. The trailing slope, instead, contains several backward (BA) like modes (i.e. modes with nodal planes in the direction perpendicular to the magnetization) with slightly higher frequency and lower intensity [19]. As the direct current is further increased towards the threshold value ($i_{DC}$=2.8 mA) the system approaches the non-linear regime and a third peak (M3), at higher frequency, emerges from the fundamental mode trailing slope and its relative intensity increases as the direct current is further increased up to a value of $i_{DC}$=3.1 mA.

3. **Micromagnetic simulations**

In order to shed light into the above complex scenario and to achieve a quantitative interpretation of the experimental data, we performed micromagnetic simulations considering the Landau-Lifshitz-Gilbert equation including the spin-orbit torque (SOT), see Eq.1. Since the injected current mainly flows into the Pt layer (more than 90%), because of its larger conductivity, the Rashba contribution is negligible and only the SOT due to the SHE has been considered to study the magnetization dynamics [22,23,24]:

$$\frac{d\mathbf{M}}{dt} = -\gamma_0 \mathbf{M} \times \mathbf{H_{EFF}} + \frac{\alpha}{M_S} \mathbf{M} \times \frac{d\mathbf{M}}{dt} - \frac{\mu_B \alpha_H J_{Pt}(x,y)}{eM_S^2 t_{Py}} \mathbf{M} \times \mathbf{M} \times \mathbf{y} \qquad (1)$$

where $\mathbf{M}$ is the magnetization and $\mathbf{H_{EFF}}$ is the effective field which takes into account the exchange field, the magnetocrystalline anisotropy, the external, the self-magnetostatic and the Oersted fields. $\alpha$, $M_S$ and $\gamma_0$ are the Gilbert damping, the saturation magnetization, and the gyromagnetic ratio respectively. $J_{Pt}(x,y)$ is the spatial distribution of the modulus of the current density in the Pt layer considering the same sign of the applied current, $\mu_B$ the Bohr Magneton, $e$ the electric charge, and $t_{Py}$ the thickness of the NiFe layer. **y** is the direction of the spin current polarization. $\alpha_H$ is the spin-Hall angle given by the ratio between the amplitude of transverse spin current density, generated in the Pt, and the charge current density flowing in it. The external magnetic field, with a modulus of 178 mT, is applied at 20° with respect the sample normal (z-axis) and with an in-plane angle of 5 degrees with respect to hard in-plane axis of the NW (y-axis). The parameters used for our numerical experiment are: exchange constant A=1.3×10$^{-11}$ J/m, spin-Hall angle $\alpha_H$=0.08, Gilbert damping 0.02, and saturation magnetization $M_s$=500×10$^3$ A/m. The discretization cell is 5×5×5 nm$^3$. The contribution of thermal fluctuations, at T=300 K, is also taken into account. [25]

Figure 3(a) shows a comparison between the experimental and simulated values of the frequencies of the modes as a function of the intensity of the injected current $i_{DC}$. We observe a good quantitative agreement between simulation and experiment in the whole range of $i_{DC}$ investigated. In particular simulations correctly reproduce the frequency evolution of the measured peaks (Fig.3(a)). When the value of $i_{DC}$ is well below the threshold ($i_{DC}$=1.0 mA) only two modes are visible in the simulated power spectrum (Fig.3(b)). The insets of Fig.3(b) show the simulated spatial profiles of the two modes: the profile of the lower frequency mode (M1) is typical of an edge-mode with the oscillations confined in a narrow region close to the edges. The profile of the higher frequency mode (M2), instead, mainly occupies the central part of the NW and reproduce the typical behavior of a fundamental mode. The frequency of both modes exhibits a red shift as a function of $i_{DC}$, being the oscillation frequency below the ferromagnetic resonance frequency estimated around 6.0 GHz. However, the frequency tunability is very small up to a critical current value where the STT from the SHE compensates the intrinsic magnetic losses due to the Gilbert damping and a self-oscillation of the magnetization is excited. This means that far from the threshold, the dynamics of the system is still in a linear regime and the two modes which appear in the spectrum are two eigenmodes of a transversely magnetized NW, with an intensity only slightly amplified by the presence of a small STT. The situation is very different when the value of the injected current becomes larger than the threshold ($i_{DC}$=3.1 mA, Fig.3(c)). Consistently with the experimental evidence, an higher frequency peak (M3) appears in the simulated power spectrum while the spatial profiles of the modes are strongly modified. First, the contrast of the modes' profiles sharpen as the contribution of the STT becomes dominant over thermal fluctuations. Moreover, the frequency gap between M1 and M2 reduces from about 1.3 GHz to 0.9 GHz and their spatial profiles are "hybridized": the edge mode (M1) becomes less localized and extends towards the center of the NW, while the oscillation of the fundamental mode (M2) extends towards the edges. A similar behavior was previously observed by Demidov et al. in a system consisting of an elliptical NiFe dot excited by a large microwave field out of the linear regime [26]. Similarly to the modes M1 and M2 also the third mode (M3) extends over the entire width of the NW, but its spatial profile looks less regular and is difficult to classify. It is clear that when the value of the injected current is increased above the threshold the system is driven through a strongly non-linear regime and the dynamics cannot be described anymore in terms of the linear eigenmodes of the system.

To deepen our comprehension of the complicated non-linear dynamics of this system above the current threshold, we also performed a wavelet transform analysis of the micromagnetic time-domain data according to the method developed by Siracusano et al. [27]. Figure 4 depicts the micromagnetic wavelet scalogram (MWS) of time traces for the magnetization $<\mathbf{m}_x(t)>$ obtained under the same configuration settings and for a value of $i_{DC}$=3.1mA. This confirms a very complex scenario with the three modes (M1, M2, M3) which exhibits an intermittent behavior characterized by a nanosecond scale switching and re-excitation, alternated with time intervals with no excitations. Such a non-stationary dynamics, characterized by sudden mode jumps, is typical of nonlinear dynamics pumped by N external mechanism. [28]

## 4. Conclusion

In conclusion we have investigated the spin wave dynamics excited by a pure spin current, generated by spin-Hall effect, in a transversely magnetized NiFe/Pt layered nanowire. A moderate amplification of the nanowire eigenmodes intensity and an almost flat frequency tunability has been observed when the intensity of the injected current is well below the threshold value of $i_{DC}$=2.8mA. As the intensity of the injected current approaches the above threshold, a large amplification of the modes intensity together with a sizeable frequency redshift has been measured. The appearance of an additional mode, at higher frequency, has been experimentally observed and quantitatively confirmed by the results of micromagnetic simulations. The results of a wavelet analysis of the simulation data in the time-frequency domain suggests the non-stationary character of the modes when the system is driven in a strongly non-linear regime. We are confident that the results of this study could stimulate further research in the field of spin-Hall driven spin wave dynamics and could affect the optimal design of future magnonic devices.

This work was supported by the European Community's Seventh Framework Programme (FP7/2007-2013) under Grant No. 318287 "LANDAUER" and by the Ministero Italiano dell'Università e della Ricerca (MIUR) under the PRIN2010 project (No. 2010ECA8P3).

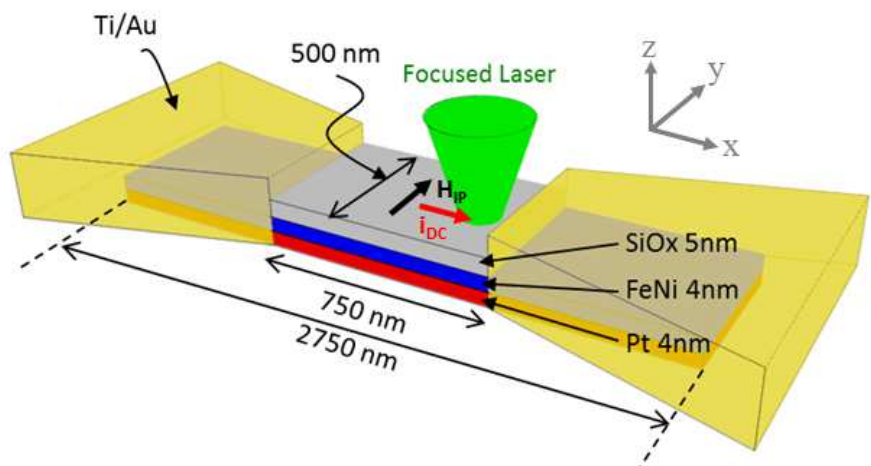

**Figure 1.** Sample layout with indication of the length, width and thickness of the different layers. The direction of the in plane component of the external field ($H_{IP}$) and of the direct current ($i_{DC}$) are also indicated.

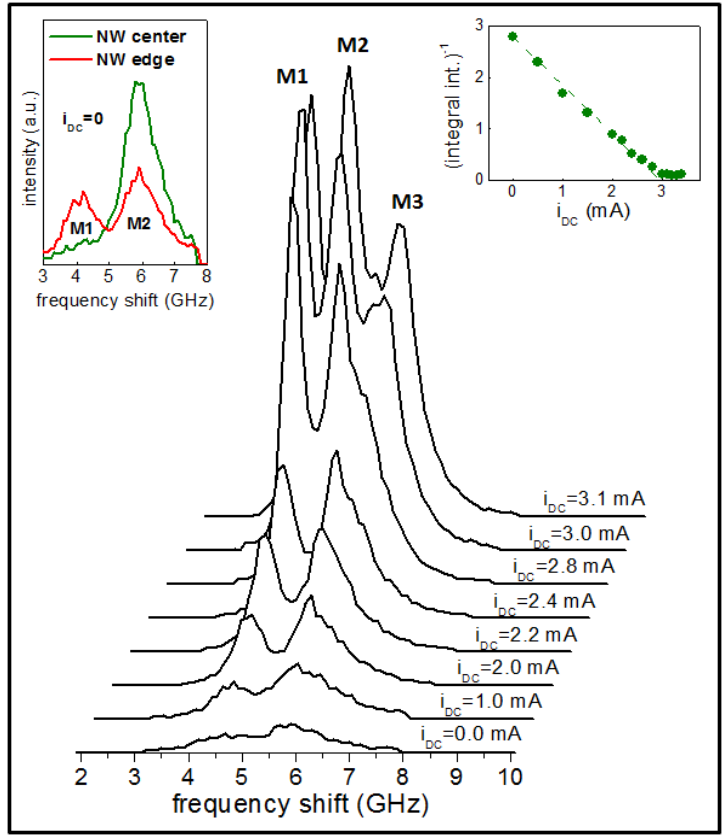

**Figure 2.** (top-left inset) micro-BLS spectra (anti-Stokes side) measured at the center and at the edge of the nanowire ($i_{DC}$=0). Sequence of micro-BLS spectra (anti-Stokes side) measured as a function of the direct current intensity $i_{DC}$. (top-right inset) inverse of the integrate intensity of the spectra as a function of $i_{DC}$.

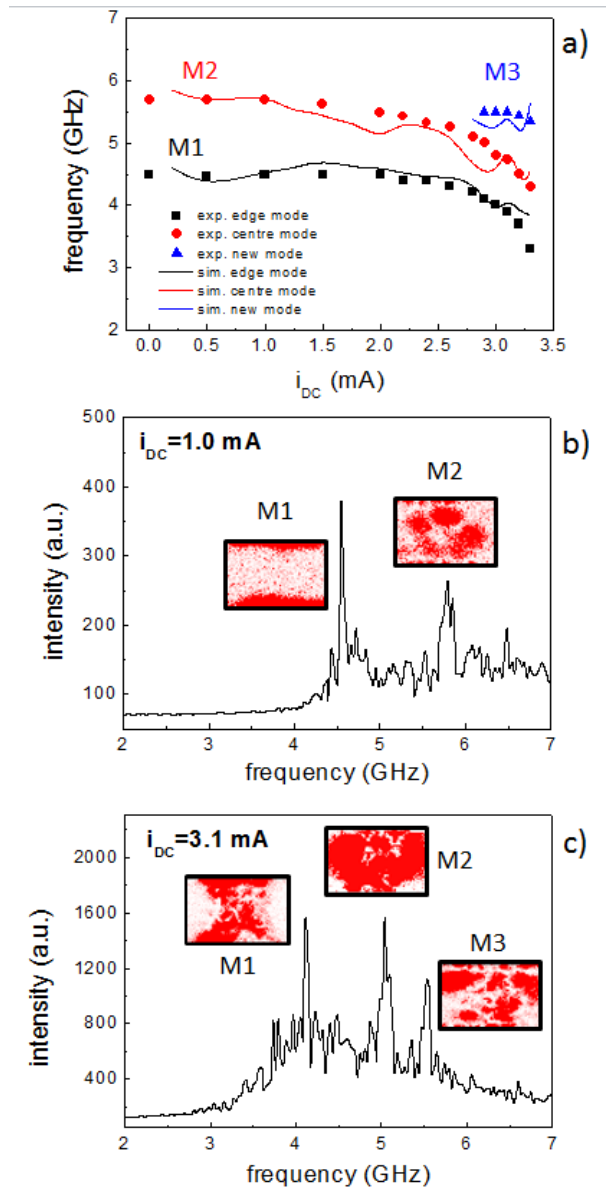

**Figure 3.** (a) Measured (points) and simulated (curve) frequencies as a function of the intensity of the direct current $i_{DC}$. (b) simulated power spectrum for $i_{DC}$=1.0 mA; (inset in b) profiles of the modes M1 and M2. (c) simulated power spectrum for $i_{DC}$=3.1 mA; (inset in c) profiles of the modes M1 and M2 and M3. The spatial profiles ($\mathbf{m}_X$ component) are calculated in the central part of the NW (750nm×500nm) where the contribution of the STT in the NiFe, due to the SHE in Pt, is larger.

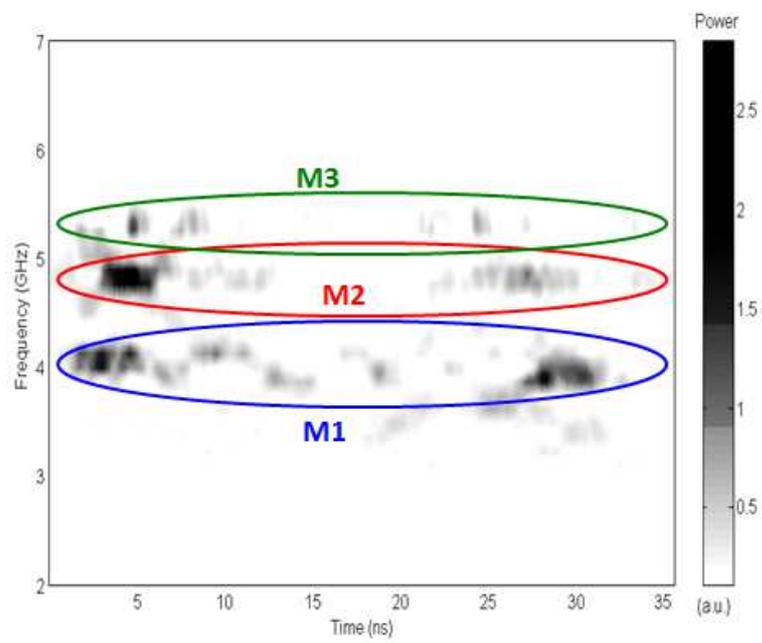

**Figure 4.** Micromagnetic wavelet scalogram (MWS) of time traces for the magnetization $<\mathbf{m}_x(t)>$ obtained for $i_{DC}$=3.1 mA.